\documentclass[twocolumn]{svjour3}          

\usepackage{graphicx}
\advance\paperheight by-20mm
\usepackage[cam,cross,horigin=-27.9mm,vorigin=-13.5mm]{crop}
\usepackage{color}
\usepackage{amsmath}
\usepackage{amssymb}
\usepackage{wasysym}
\usepackage{color,soul}
\usepackage{microtype}

\usepackage{marvosym}
\newcommand{\envelope}{(\raisebox{-.5pt}{\scalebox{1.45}{\Letter}}\kern-1.7pt)}

\usepackage{times}
\usepackage[LY1]{fontenc}
\usepackage{caption}
\usepackage{subcaption}
\usepackage{booktabs}
\usepackage{todo}

\begin{document}

\title{A Single-Ion Trap with Minimized Ion-Environment Interactions }

\author{P.B.R.\, Nisbet-Jones \and S.A.\,King \and J.M.\,Jones \and R.M.\,Godun \and C.F.A\,Baynham \and K.\,Bongs \and M.\,Dole\v{z}al \and P.\,Balling \and P.\,Gill}

\date{Received: 18 October 2015}

\institute{P.B.R.\, Nisbet-Jones \and S.A.\,King \and J.M.\,Jones \and R.M.\,Godun  \and C.\,Baynham \and P.\,Gill \at National Physical Laboratory, Hampton Rd, Teddington, Middlesex, TW11 0LW, UK \\
\email{rachel.godun@npl.co.uk} \and M.\,Dole\v{z}al \and P.\,Balling \at Czech Metrology Institute, V Botanice, 150 72, Prague, CZ \and J.M.\,Jones \and K.\,Bongs \at
School of Physics and Astronomy, University of Birmingham, Birmingham, B15 2TT, UK \columncase{,}{\\}}

\maketitle

\begin{abstract}
We present a new single-ion endcap trap for high precision spectroscopy that has been designed to minimize ion-environment interactions.  We describe the design in detail and then characterize the working trap using a single trapped $^{171}$Yb$^{+}$ ion.  Excess micromotion has been eliminated to the resolution of the detection method and the trap exhibits an anomalous phonon heating rate of $d\langle n\rangle /dt = 35 \pm 20$\,s$^{-1}$.  The thermal properties of the trap structure have also been measured with an effective temperature rise at the ion's position of $\Delta T_{\rm{(ion)}} = 0.14 \pm 0.14$\,K. The small perturbations to the ion caused by this trap make it suitable to be used for an optical frequency standard with fractional uncertainties below the $10^{-18}$ level.

\end{abstract}

\section{Introduction}

Trapped single atomic ions can provide a near-idealized system in which a quantum object is decoupled from its environment, and any interactions can be precisely tailored.  Many uses for such traps have been found, from frequency metrology and tests of fundamental physics \cite{Godun2014a,Huntemann2014a,Wansbeek2008}, to simulations of quantum systems \cite{Gerritsma2010,Menicucci2007} and quantum information processing \cite{Hucul2015,Gessner2013}, however the imperfections inherent with any physical system introduce perturbations and uncertainties to the experiments being performed. In many cases these can limit the operational capability of the trapping system. 

This paper presents a new single-ion endcap trap specifically designed to reduce the ion's interaction with its environment.   Particular care has been taken to minimize dc and rf phase-offset induced micromotion, thermal effects, and ion heating rates, and also to maximize optical access, resulting in significant improvements over existing designs.  Although this trap was developed with a particular interest in precision spectroscopy of the $^{2}$S$_{1/2}-^{2}$F$_{7/2}$ electric octupole (E3) transition in $^{171}$Yb$^{+}$ for frequency metrology and tests of fundamental physics, its performance is applicable to ion trapping for a wide variety of applications.   

The requirements on an ion trap for fundamental frequency metrology with fractional uncertainty at the $10^{-18}$ level are presented in Section\,\ref{sec:req}.  Details of the design and fabrication of the trap which meets these requirements are given in Section\,\ref{sec:design}. Characterization of the trap using a single $^{171}$Yb$^{+}$ ion is performed in Section\,\ref{sec:charac}. 

\section{Trap Requirements}\label{sec:req}

An ion trap and vacuum system for high precision frequency spectroscopy must:
\begin{enumerate} 
\item{Minimize the thermal black-body radiation (BBR) emissions from the trap, resulting in the BBR environment at the ion's position being dominated by the well controlled ambient field;}
\item{Minimize the anomalous phonon heating rate, $\Gamma_{\rm{h}}$ of the ion;}
\item{Minimize stray electric fields which result in rf-induced micromotion;}
\item{Minimize collisions with background gas;}
\item{Maximize fluorescence collection from the ion to improve the state detection signal-to-noise ratio (SNR).}
\end{enumerate}

The first two of these requirements are in conflict.  The dominant thermal heating mechanism of the ion trap structure is dielectric loss in the insulators which isolate the rf and ground sections of the trap.  This loss takes the form $P = V\epsilon'_{r}\epsilon_{0}\tan\delta\Omega_{\rm{rf}} E^{2}$, where the power $P$ dissipated into a dielectric of volume $V$ depends on the dielectric's relative permittivity $\epsilon'_{r}$ and loss tangent $\tan\delta$ when an electric field $E$ of frequency $\Omega_{\rm{rf}}$ is applied across it. Whilst selecting materials which have a low loss tangent and relative permittivity is important, the easiest way to reduce the thermal heating of the trap structure is to operate the trap with the lowest possible values for $\Omega_{\rm{rf}}$ and $E$.  However, to achieve a low phonon heating rate $\Gamma_{\rm{h}}$ the trap should be operated at the highest possible drive frequency and electric field to provide tight ion confinement and hence high secular frequencies $\omega_{r,z}$.

The anomalous heating rate also has a $z_{0}^{-\alpha}$ dependence on the ion-electrode separation $z_{0}$. The exponent $\alpha$ depends on the noise process involved but comparison between traps produced in different laboratories indicate a range of $2<$\,$\alpha$\,$<4$ \cite{Brownnutt2014} making it highly beneficial to have a large electrode spacing. To maintain a given secular frequency $\omega_{r,z}$ when changing $z_{0}$, $E$ must be increased as $z_{0}^{2}$. These effects make it clear that careful design is required to find a suitable compromise configuration in which the trap can operate successfully within the above constraints. 

The other requirements are more straightforward to meet.  Relative phase delays between the rf voltage on the two endcaps will result in micromotion that cannot be compensated using only dc fields.   The trap structure must therefore be totally symmetric to ensure that the effective path length between the endcaps and the rf feed, including all possible capacitive and inductive delays, is equal.  Likewise the ion must be shielded from any rf field that does not come from the endcaps. The ion trap must be contained within an UHV/XHV environment to reduce pressure-dependent shifts on the order of $\sim10^5$\,Hz\,Pa$^{-1}$ \cite{Dube2013}. Finally, any apertures in the imaging setup should be placed as close to the ion as possible to increase the numerical aperture (NA) and hence fluorescence collection efficiency. Fast state detection is necessary for transitions with short excited state lifetimes to avoid losing excitation probability \cite{Peik2006}, increasing the clock instability via increased dead time, and introducing the Dick effect from the resulting under-sampling of the local oscillator noise.   A high SNR also allows easy use of photon correlation spectroscopy to minimize the excess micromotion. 

\section{Trap Design and Fabrication}\label{sec:design}

The trap presented in this paper is based on the endcap design \cite{Schrama1993}, and consists of two cylindrical electrodes that are held at an equal rf potential facing one another whilst contained within a pair of independent dc electrodes. This basic design is in use in multiple laboratories, \cite{Godun2014a,Madej2012,Takahashi2013} and has the benefit of excellent optical access in the radial plane. Recent experiments with fiber-optics contained within the endcap electrodes have extended this access into the axial direction \cite{Takahashi2013}.

\begin{figure}[hbtp]
\begin{center}
\includegraphics[width=\columnwidth]{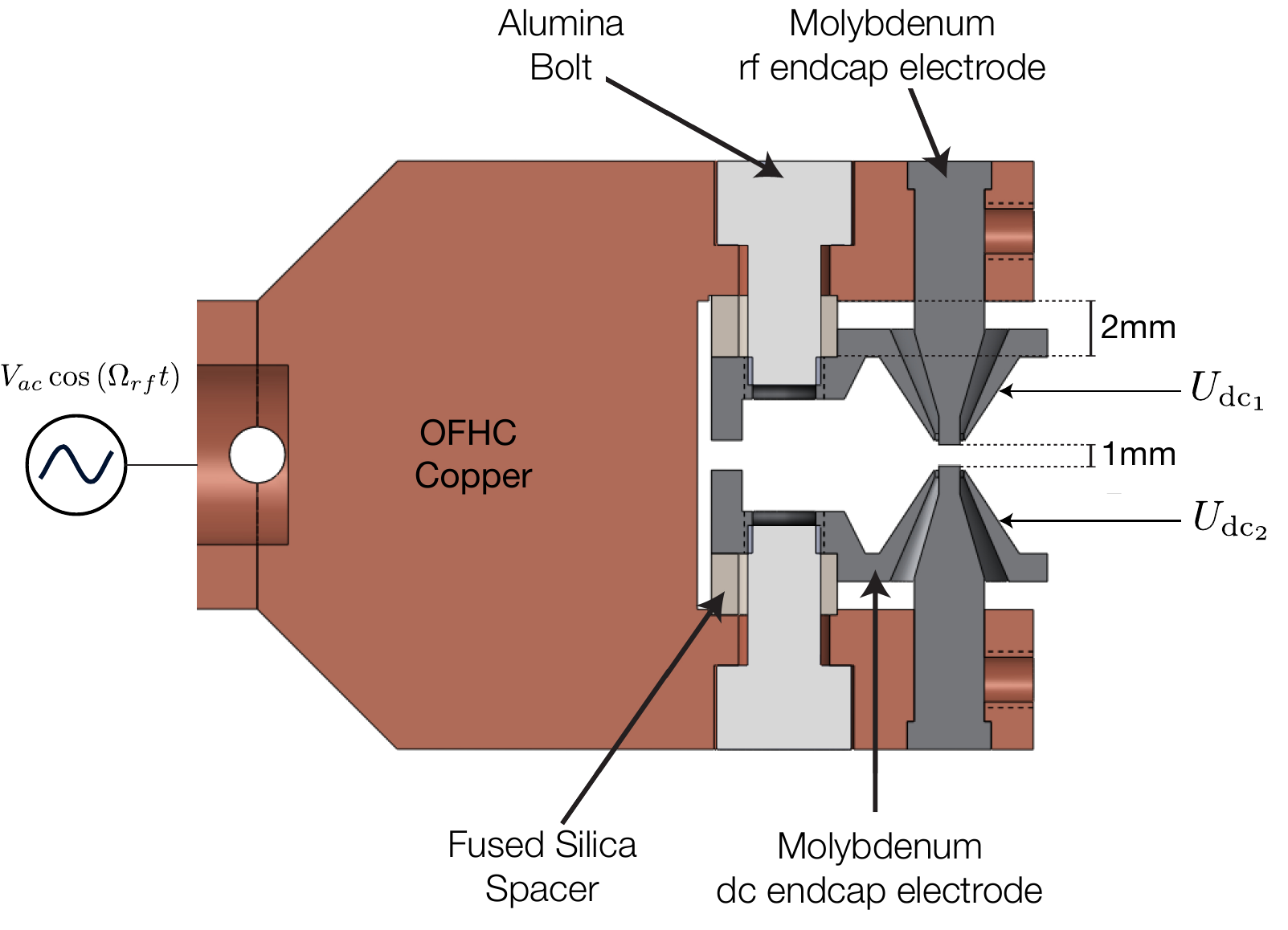}
\caption{Vertical cross-section through the trap structure.  The rf endcaps are secured into the OFHC copper with M2 set-screws and sit within hollow dc endcap electrodes that are supported by fused silica spacers and recessed alumina bolts.}
\label{fig:trap}
\end{center}
\end{figure}

\begin{figure}[htbp]
\begin{center}
\includegraphics[width=\columnwidth]{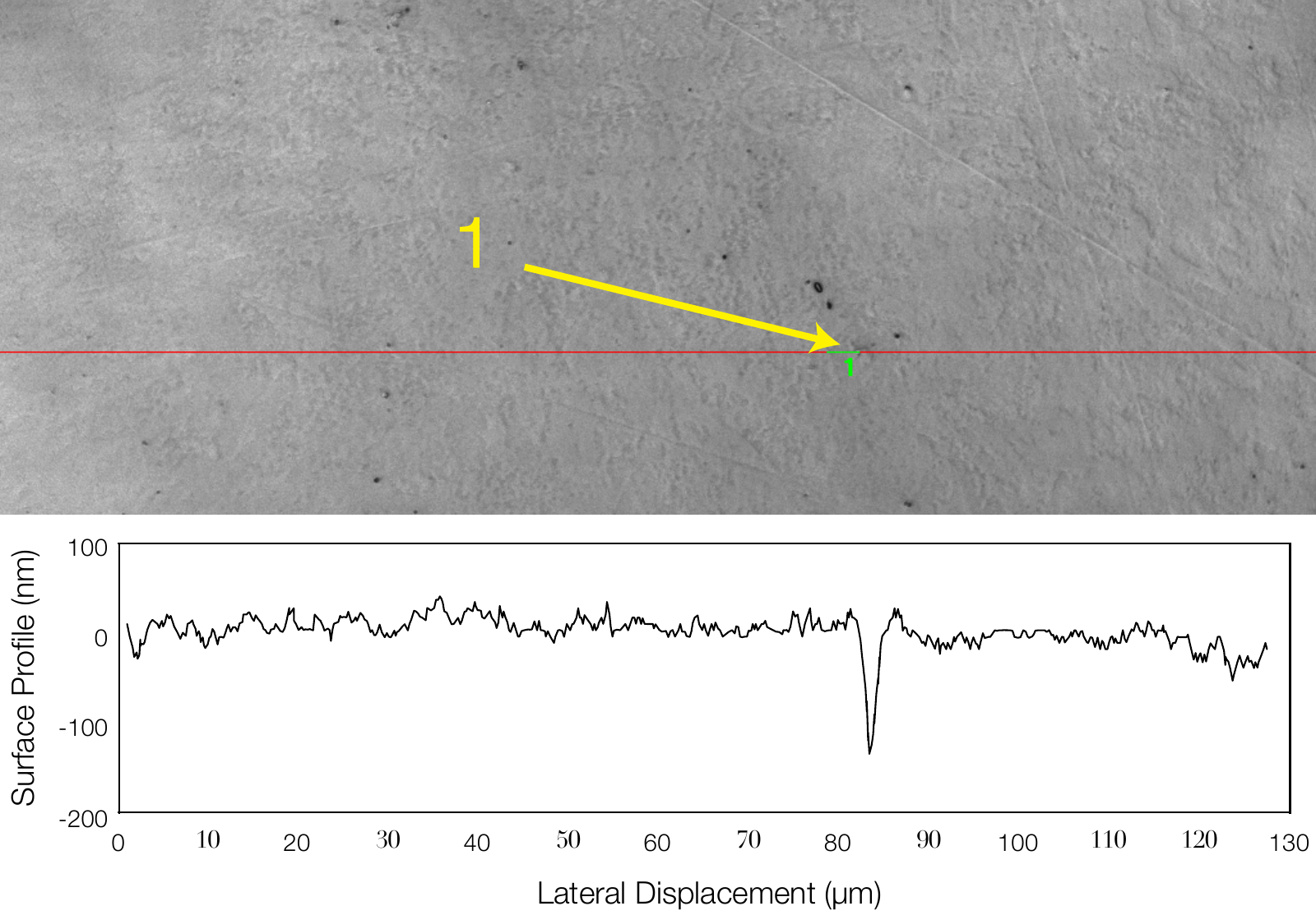}
\caption{Focal variation microscopy image of the polished surface of the upper rf endcap electrode.  The inset plot shows the surface profile along the red line.  The 150\,nm pit highlighted by the yellow marker (1) is typical of the other imperfections which are visible as black dots in this image, although on a larger scale. }
\label{fig:profile}
\end{center}
\end{figure}

A schematic of the trap is shown in Figure\,\ref{fig:trap}.  The support structure for the trap, which also acts as the high voltage rf feed, is made of oxygen-free, high thermal conductivity (OFHC) copper.  OFHC copper has both high electrical and thermal conductivity, and exhibits low out-gassing which will reduce ion loss via chemical reactions of the ion with oxygen. The two 99.9\% purity molybdenum rf endcap electrodes  are secured into the copper mount using an interference fit and M2 set-screws. The electrode tips have a diameter of $750\mu$m and a separation of $1000\mu$m. Molybdenum has good electrical and thermal conductivity, is non-magnetic, and can be polished to a high surface quality.  Material properties are also thought to contribute to $\Gamma_{\rm{h}}$ and whilst the mechanism behind this is unclear, traps with cleaned Mo electrodes have demonstrated excellent results \cite{Turchette2000}.  The faces of the rf electrodes which subtend the ion were mechanically polished to achieve a mirror quality finish as shown in Figure\,\ref{fig:profile}.  The mean surface roughness was measured via focal-variation microscopy to be $\rm{R}_{\rm{a}}=20$\,nm, resulting in an emissivity of $\epsilon\approx0.02$ at 300\,K \cite{Ordal1988}.   As the inner endcap electrodes represent the largest solid angle contribution to the ion's field of view, the low emissivity reduces the thermal radiation that the ion will experience from the temperature of these endcaps.  The polished surface also has the potential to reduce $\Gamma_{\rm{h}}$ as it removes any possible points for field emission of electrons. Mechanical polishing also exposes a raw surface free from adsorbates which have been proposed as a source of electric field noise \cite{Brownnutt2014}.    Although in-vacuum cleaning to remove adsorbates was not attempted \cite{Hite2012}, the clean electrodes were kept in a protective environment to minimize re-contamination of the surface before UHV conditions were achieved.

Two molybdenum dc electrodes surround the rf endcaps and form the ground reference for the rf potential in the trapping region. The separation between the rf and dc electrodes takes a conical cross-section. A small separation of 135\,$\mu$m near the trapping region ensures a high quadrupole efficiency  of $\varepsilon=0.67$ \cite{Schrama1993}, expanding to 1\,mm at the base to reduce the capacitance between the endcaps, and thus reduce the rf current required to drive the trap.  This also allows the rf electrodes to increase in thickness near the base increasing their mechanical rigidity.  Were the large separation to be maintained, the trap efficiency would fall by $\sim$50\,\% in addition to the degraded optical access.  The dc electrodes are positioned using a pair of fused silica spacers and are secured with 99.9\% purity alumina bolts.  By placing the dielectrics in a non-critical region away from the trap center, it is possible to have a large separation between the high voltage and dc electrodes greatly reducing $E^{2}$ across the lossy dielectric.    The position of the high emissivity dielectric behind the dc endcap also removes any direct line of sight between it and the ion.  

\begin{figure}[htbp]
	\centering
	\begin{subfigure}[b]{0.45\columnwidth}	
		\includegraphics[width=\columnwidth]{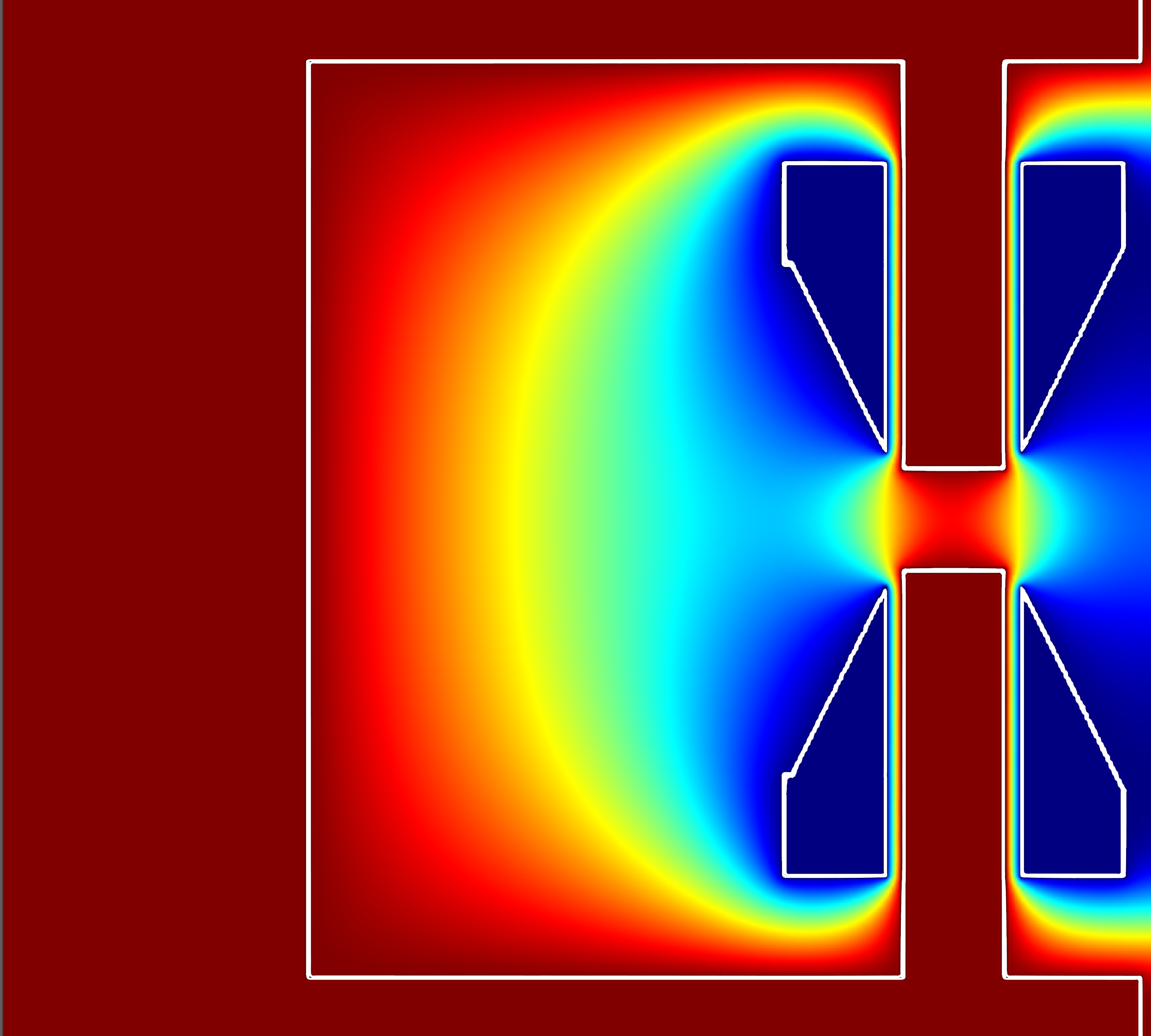}
	\end{subfigure}
	\begin{subfigure}[b]{0.45\columnwidth}	
		\includegraphics[width=\columnwidth]{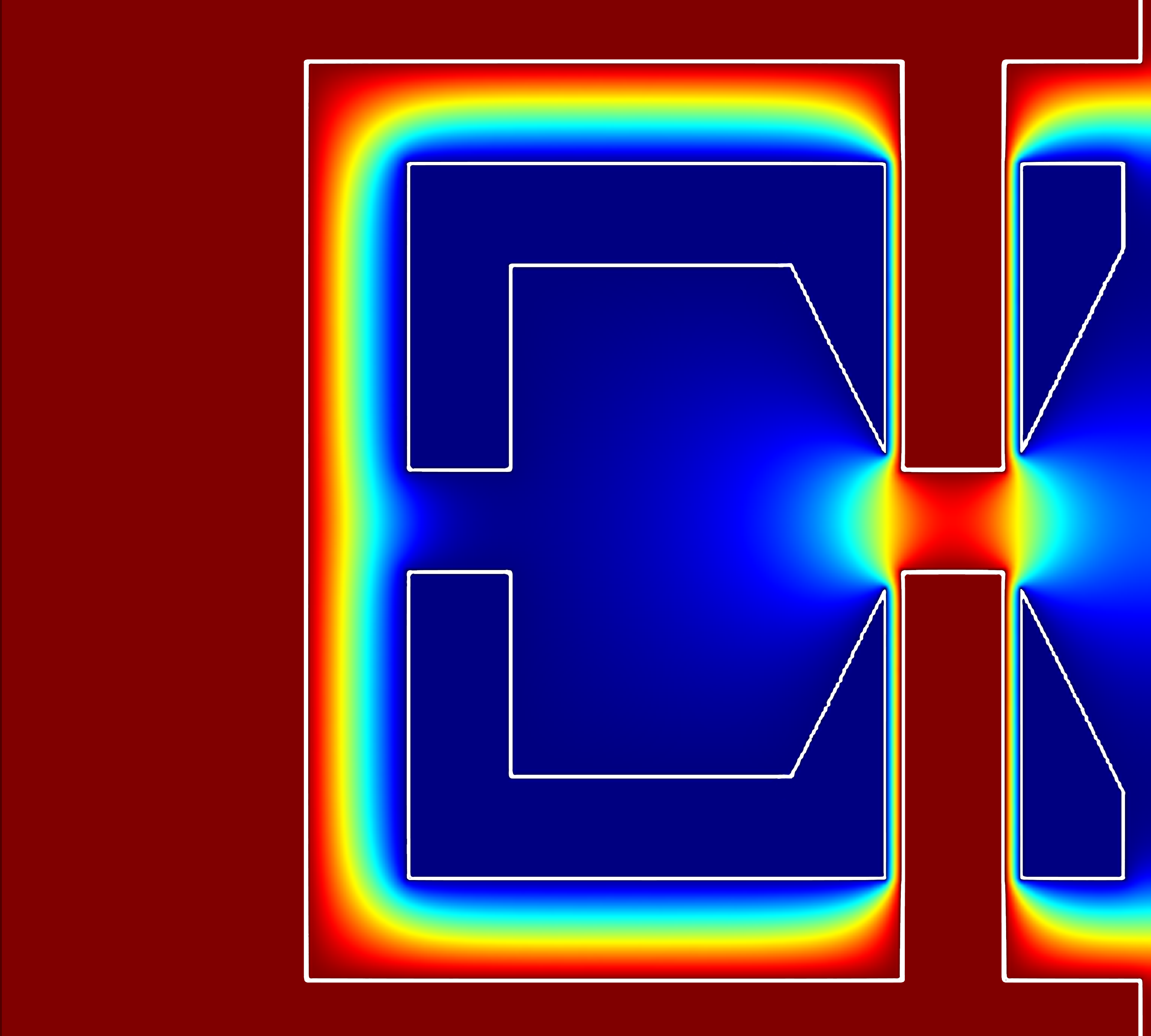}
	\end{subfigure}

	\caption{Illustration of the importance of extending the dc electrodes to shield the trapping region from the rf potential caused by the C-shaped mount.   Simulation of the rf electric potential performed using the COMSOL software package.  With the copper C-shaped mount unshielded the rf potential asymmetrically distorts the quadrupole potential in the trapping region.}
	\label{fig:comsol}
\end{figure}

The C-shape of the copper mount breaks the trap's cylindrical symmetry. The asymmetric mounting protrusions on the dc endcaps effectively shield the ion from the copper mount restoring the surrounding ground potential (Figure\,\ref{fig:comsol}). Without the extra shielding the rf minimum would be shifted from the geometric center of the trap, and there would no longer be a true micromotion zero due to the phase delay between the rf potential from the electrode tips and the center of the C.

The dc electric field in the trapping region is controlled by applying static voltages to the dc electrodes (axial direction) and to two fixed compensation electrodes set back from trap center by ~5\,mm in the radial plane (Figure\,\ref{fig:Cartoon}) with a separation of 60$^{\circ}$ to maximize optical access in the radial plane.  The dc electrodes are capacitively shorted to ground via 1\,$\mu$F capacitors \footnote{High SRF SMD components (X7R dielectric) ensures good performance in the 1-10\,MHz range.} to prevent any rf pickup on the dc electrodes that could be re-radiated with a phase offset. These capacitors are located 50\,mm from trap center on the air-side of the dc feedthrough.  To avoid phonon heating from voltage noise on the compensation electrodes, the applied dc voltages are filtered close to the feedthrough using an RC filter followed by a CLC $\pi$-filter, providing 70\,dB of voltage noise attenuation at 1\,MHz.\footnote{R\,=\,100\,k$\Omega$, C\,=\,100\,nF, L\,=\,22\,$\mu$H (Coilcraft 1812CS)}$^{,}$\footnote{poles at 10\,Hz and 100\,kHz respectively.}

Isotopically enriched neutral $^{171}$Yb is contained within a resistively heated tantalum tube oven.  The atomic beam is slightly collimated (opening angle of $\theta\approx10^{\circ}$) reducing the amount of Yb deposited on the trap structure during loading. Yb deposits can cause patch potentials leading to changes in the necessary compensation voltages and might also affect the phonon heating rate of the ion.

\subsection {Vacuum System}\label{sec:vac}

\begin{figure*}[htbp]
\begin{center}
\includegraphics[width=2\columnwidth]{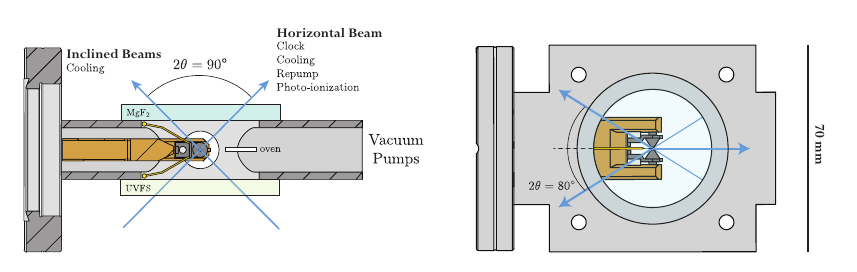}
\caption{Schematic diagram of the trap and vacuum chamber. (a) Cross-section in the horizontal plane taken through the trap center. (b) Front view.  The compensation electrodes are shown in yellow and the paths of the named co-linear laser beams are shown as blue lines.} 
\label{fig:Cartoon}
\end{center}
\end{figure*}

The vacuum system maximizes the optical access to the trap whilst maintaining a small light-weight package. The chamber was machined out of a single aluminum block and is fitted with non-magnetic CF mounting flanges.   The trap region of the chamber takes the form of a $70\times70\times20$\,mm cuboid with a central $\diameter=40$\,mm cylindrical opening (Figure\,\ref{fig:Cartoon}).  This opening is sealed using two $\diameter=50$\,mm viewports. The laser access and ion imaging window is anti-reflection coated UV-grade fused silica (UVFS).  The other window is magnesium fluoride (MgF$_{2}$), which is transparent up to 8\,$\mu$m to allow direct thermal imaging of the trap structure during operation.  Both windows are cold-welded to the chamber with $\diameter=1$\,mm indium wire, achieving undetectable He leak rates.

The indium sealed windows allow a maximum optical access opening angle of $2\theta=112\,^{\circ}$, compared to the $2\theta=80\,^{\circ}$ from a traditional CF optic, and allows us to probe and cool the ion in two perpendicular radial directions.  This also maximizes the effectiveness of the polarization spinning technique  \cite{Berkeland2001} and ensures a high fluorescence rate. Optical access in the axial direction is limited by the trap structure itself at $2\theta=103\,^{\circ}$.

Ultra-high vacuum is maintained by a SAES NexTorr D100-5 combination non-evaporable getter (NEG) and ion-getter pump. The NEG provides a pumping speed of up to 100\,ls$^{-1}$ for active gases (H$_{2}$, H$_{2}$O, N$_{2}$, O$_{2}$, etc.), while inert gases are pumped by the ion getter pump at a rate of 6\,ls$^{-1}$.  After final assembly a bakeout at 125$^{\circ}$C roughed by a turbo-molecular pump achieved a base pressure of $<1\times10^{-9}$\,Pa ($10^{-11}$\,mbar) determined by the current draw of the ion pump.  For a Langevin-approximated frequency shift of 10$^5$\,Hz\,Pa$^{-1}$ \cite{Dube2013} this corresponds to 100\,$\mu$Hz. 

\subsection{Light delivery and collection}

The ion is illuminated along three optical axes which probe perpendicular radial directions and sample the axial motion of the ion at an angle of $\theta=40\,^{\circ}$ (Figure\,\ref{fig:Cartoon}).  The horizontal beam contains all the wavelengths that are used in the experiment which range from the UV (369\,nm cooling) to the IR (935\,nm re-pump).  These wavelengths are combined using dichroic beamsplitters and coupled into an endlessly-single-mode photonic crystal fibre\,\footnote{NKT LMA-PM-5}.  The fibre output is brought to an achromatic focus of waist $w=19\,\mu$m at the ion's position using a pair of off-axis parabolic mirrors, ensuring perfect overlap of all lasers onto the ion.  The two beams that are raised out of the horizontal plane only supply cooling light.

Atomic fluorescence is collected through the UVFS viewport using an aspheric lens pair in a 2.6:1 telescope. After spatial and spectral filtering\,\footnote{NA = 0.40, Filter Transmission T=92\,\%, PMT Q.E. = 27\,\%} an on-resonance fluorescence signal of $20,000\,$s$^{-1}$ is obtained for a single ion at $I_{\rm{sat}}$.  For a typical background count rate of $50\,\rm{s}^{-1}$ this results in a signal to noise ratio of SNR = 400.

\section{Trap Characterization}\label{sec:charac}

The trap performance was characterized using single $^{171}$Yb$^{+}$ ions however the performance should be directly transferable to other elements.  Ions have been loaded and trapped at rf frequencies from $\Omega_{\rm{rf}}/2\pi=$\,$3$\,$-$\,$21$\,MHz.  The secular frequencies of trapped single ions were measured by `tickling' the ion with a low-power rf field applied to one of the dc endcaps. When resonant with the ion's secular frequency this tickling field heats the ion, and greatly increases the observed fluorescence from a far red-detuned cooling laser.   Secular frequencies have been arbitrarily set from the kHz to the MHz range up to a maximum of $\omega_{(x,y,z)}=2\pi\times(1.02,1.04,2.04)$\,MHz, obtained using a helical resonator with a loaded resonant frequency of $14$\,MHz with 1\,W of forward power. Ion lifetimes on the order of a week are routinely observed at all frequencies.  Uncooled ion lifetimes of up to 15\,hrs have been observed on multiple occasions.  Collisions with background gas molecules can populate the long-lived F-state in Yb$^{+}$,  and without repumping, is equivalent to loss of the ion. We routinely observe $>24$\,hr lifetimes without collisional transfer to this state which supports the low vacuum pressure measured in Section\,\ref{sec:vac}.

\subsection{3D Micromotion minimization}\label{sec:micromotion}

For ion-based atomic clocks, residual excess micromotion from imperfectly compensated stray potentials causes two of the most significant trap-induced systematic shifts: the dc-Stark shift and the time-dilation shift \cite{Godun2014a,Huntemann2012,Chou2010a}. Additionally, when pushed away from the rf potential minimum the ion is exposed to the possibility of electrical noise from the trap drive.

The projection of the ion's micromotion in the direction of any one of the cooling beams can be measured using the photon correlation technique.   The excess micromotion induces a first-order Doppler shift of the cooling laser in the frame of the ion, which can be observed as a modulation of the ion's fluorescence at the trap drive frequency when the ion is illuminated with light at the cooling transition half-maximum.  The observed fluorescence signal $S_{\rm{obs}}$ from each cooling beam can then be written as 
\begin{equation}
S_{\rm{obs}} = \frac{S_{\rm{max}}}{2}\left[ 1 + S_{\rm{mod}}\sin(\Omega_{\rm{rf}}t+\phi)\right],
\end{equation}
where the maximum signal $S_{\rm{max}}$ is sinusoidally modulated at the trap drive frequency $\Omega_{\rm{rf}}$ with phase $\phi$ and fractional amplitude $S_{\rm{mod}}$.  To minimize the micromotion, voltages are applied to the compensation electrodes and the resulting electric field pushes the ion towards the rf potential minimum reducing the value of $S_{\rm{mod}}$.  A complete analysis of this technique, including effects caused by the fluorescence transition's finite linewidth is given in \cite{Keller2015}.

\begin{figure}[hbtp]
\begin{center}
\includegraphics[width=\columnwidth]{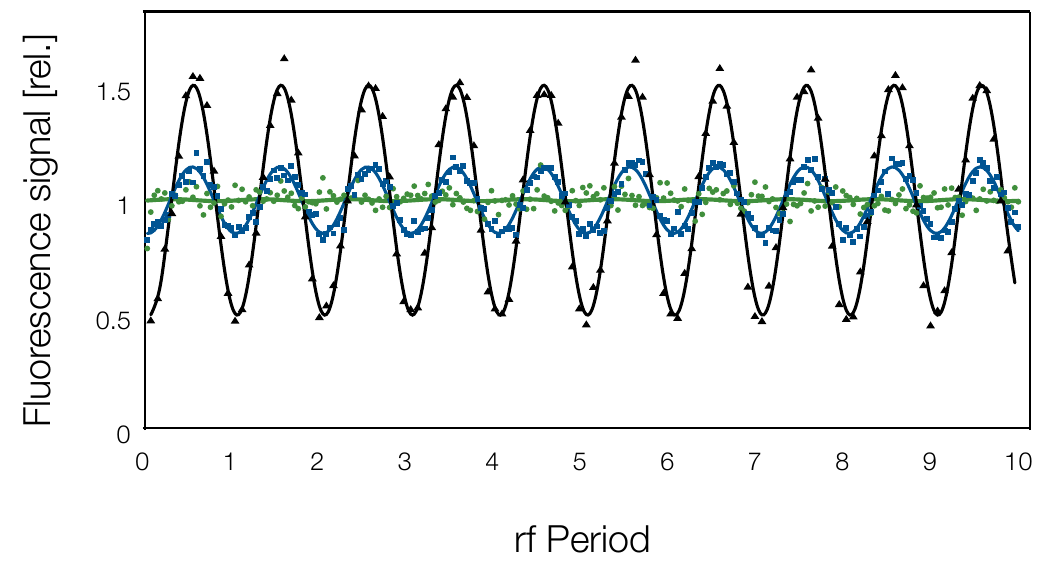}
\caption{Photon correlation signal and sinusoidal fits for an ion with a fluorescence modulation of $S_{\rm{mod}}=0.5$ (black triangles), $S_{\rm{mod}}=0.12$ (blue squares), and $S_{\rm{mod}}=0.01$ (green circles).      }
\label{fig:RFphoton}
\end{center}
\end{figure}

\begin{figure}[htbp]
\begin{center}
\includegraphics[width=\columnwidth]{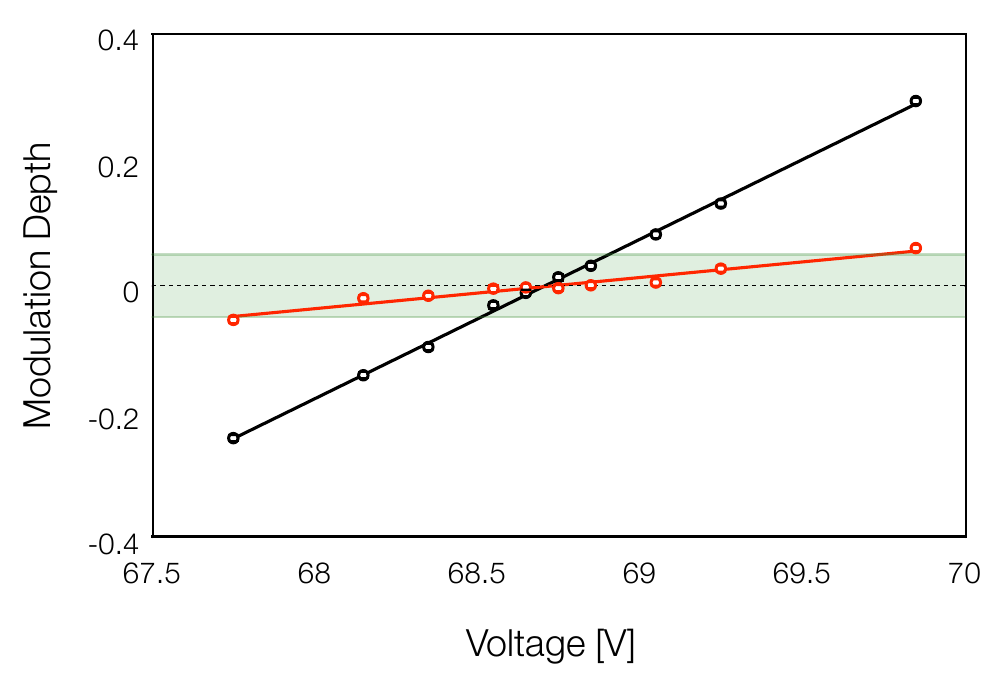}
\caption{Modulation depth of the photon correlation signal as a function of the voltage on one of the compensation electrodes in the radial plane for the beams that are at an angle of $75^{\circ}$ (black)  or $15^{\circ}$ (red) in the radial plane.  The green shaded area represents the modulation depth where the dc-Stark and time dilation shifts are below $1\times10^{-18}$.  A sign change in the modulation depth corresponds to a $\pi$ phase change in the modulation signal.}
\label{fig:C2}
\end{center}
\end{figure}

An example of the resulting correlation plot is shown in Figure\,\ref{fig:RFphoton} for three different levels of micromotion.    Micromotion along the two perpendicular axes of the beams in the radial plane is shown in Figure\,\ref{fig:C2} as a function of the voltage applied to one of the radial compensation electrodes. As the laser beams have components along both of the ion-to-compensation-electrode axes, changing one of the compensation voltages will alter the micromotion observed along both beams.  The micromotion projection along a given beam is minimized using the more sensitive compensation electrode, and through an iterative process can be eliminated to the measurement resolution of the system, resulting in $S_{\rm{mod}}=0.000(5)$.
          
 After micromotion minimization the trap has been consecutively reloaded up to twenty times over a 5\,hour period without noticeable changes in the micromotion at the $S_{\rm{mod}}=0.01$ level being observed.  No long term drifts in the micromotion have been observed suggesting that vacuum windows are sufficiently far from the ion that dielectric charging from the UV laser beams is insignificant \cite{Harlander2010} and that the trap is electrically and mechanically stable.

\subsection{Anomalous heating rate of the ion}

In many experiments such as optical atomic clocks or quantum computation based on optical qubits, it is desirable to resolve narrow lineshapes on electric dipole forbidden transitions.  The peak excitation (gate fidelity) is limited by the dephasing caused by the thermal distribution of the vibrational quantum number of the ion. Even if an ion is cooled to its vibrational ground state before interrogation, electric field noise will cause the ion to heat during the probe pulse, thus reducing the fringe contrast.      As the wavefunction of a hot ion is larger and it experiences more of the trapping potential a high heating rate will also increase the dc Stark and time-dilation shifts caused by the ion's secular motion.

The phonon heating rate of the trap was measured using the Doppler re-cooling technique.  After a prolonged period without laser cooling the cooling laser, tuned close to the unperturbed line center, is re-applied to the ion. The temperature of the ion before re-application of the cooling laser can be determined by analysis of how the fluorescence returns.  A full description of this technique can be found in \cite{Wesenberg2007,Epstein2007,Brownnutt2014}.

\begin{figure}[tbph]
	\centering

	\begin{subfigure}[b]{\columnwidth}	
		\includegraphics[width=\columnwidth]{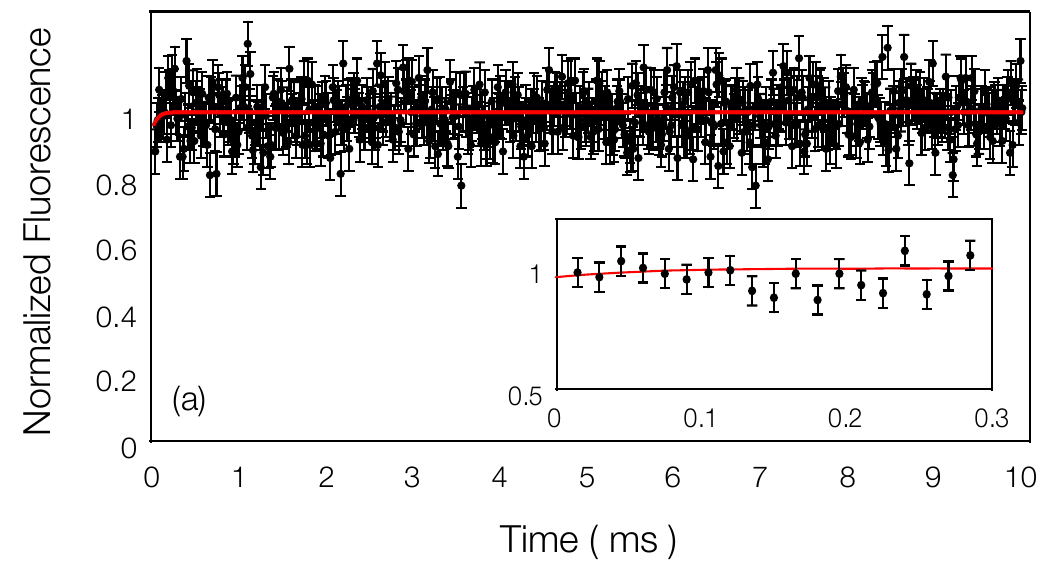}
	\end{subfigure}

	\begin{subfigure}[b]{\columnwidth}	
		\includegraphics[width=\columnwidth]{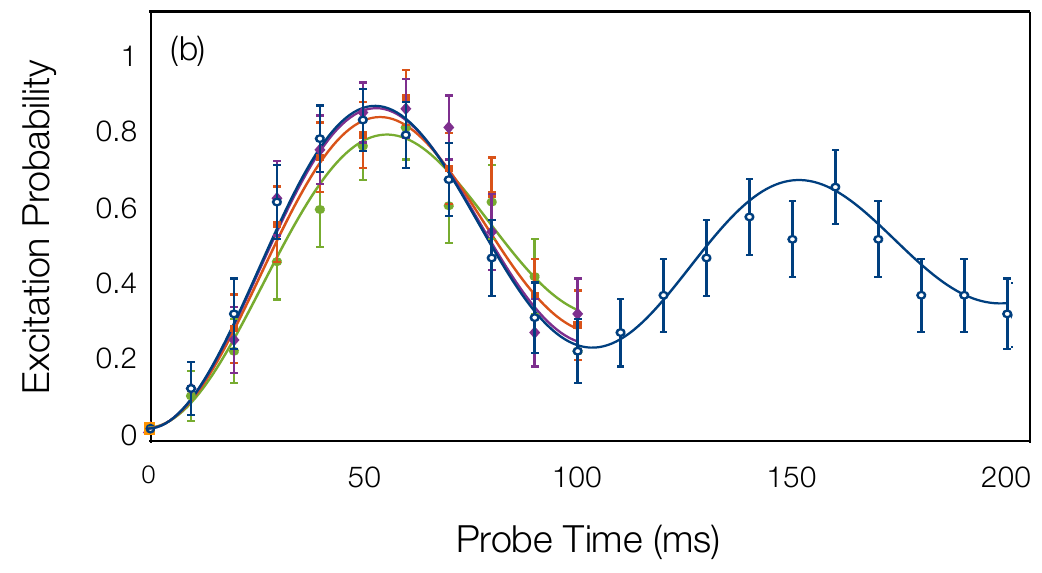}
	\end{subfigure}

	\caption{The top plot shows ion fluorescence as a function of re-cooling time for a Doppler re-cooling measurement of the trap anomalous heating rate. The ion was illuminated at $I=I_{\rm{sat}}$ and at a detuning of 5\,MHz. Fluorescence return  was measured after 30\,s without cooling.  The red line shows the theoretical fluorescence return for a heating rate $d\langle n\rangle /dt = 50$.  The inset plot shows the first 300\,$\mu$s of recooling. The bottom plot shows the Rabi oscillation decay on the E3 transition for post-cooling delay times of 10\,ms (hollow circles), 50\,ms (purple diamonds), 200\,ms (orange squares), and 500\,ms (green circles).  Peak excitation of the first Rabi pulse was limited to 85\% by noise on the excitation laser.}
	\label{fig:phonon}
\end{figure}

The trap was operated at an rf frequency of $\Omega_{\rm{rf}}/2\pi = 7.62$\,MHz with a secular frequency $\omega_{r}/2\pi=500$\,kHz.  This is much lower than the natural linewidth of the cooling transition $\Gamma/2\pi = 19.6$\,MHz, ensuring that the ion is within the weak-binding limit where the semi-classical Doppler cooling model is valid.   
The ion was allowed to heat for periods up to 50\,s before the 5\,MHz red-detuned cooling laser was reapplied and the fluorescence return counted in 15\,$\mu$s time-bins. The fluorescence return was measured 3000 times to decrease the statistical uncertainty and the results averaged as shown in Figure\,\ref{fig:phonon}(a).   There was no resolvable change in the fluorescence level throughout the 10\,ms counting window. The resolution of the measurement was limited to $d\langle n\rangle /dt = 100$\,s$^{-1}$ by the signal-to-noise achieved in our 15\,$\mu$s binning of the photon counter signal.  This yields an upper limit on the phonon heating rate of $d\langle n\rangle /dt = 50 ^{+100}_{-50}$\,s$^{-1}$ with $1\sigma$ uncertainty curtailed at zero.

An alternative method of measuring the anomalous heating rate is to observe dephasing of the Rabi oscillations on an atomic transition as a function of the time delay between cooling the ion and starting a Rabi transition. The excitation probability $P$ can be calculated from the excitation probability $P_{\rm{n}}$  and Rabi frequency $\Omega_n$ for each motional state $n$ \cite{Letchumanan2004} along with the probe-laser-induced decoherence over the probe time $t$, caused by its  finite coherence time $\tau$, and occupation probability of the clock ground state $A$:

\begin{align}
P_{}=\frac{A}{2}\left(1-e^{-t/\tau}\sum_{n}P_{\rm{n}}\cos\left(\Omega_{n} t \right) \right)
\label{eq:rabi}
\end{align}

Ion heating during the probe pulse can also affect the dephasing rate, however for the initial temperature corresponding to $\langle n\rangle \sim 30 $, and the comparatively short probe time, this effect can be neglected. During this measurement the trap was operated at an rf frequency of $\Omega_{\rm{rf}}/2\pi = 20.89$\,MHz with a secular frequency $\omega_{r}/2\pi=776$\,kHz. Rabi oscillations were driven on the 467\,nm: $^{2}$S$_{1/2}-^{2}$F$_{7/2}$ transition following a delay of up to 500\,ms between the end of the cooling cycle and interrogating the transition. Fitting Equation\,\ref{eq:rabi} to this data (Fig.\,\ref{fig:phonon}b) determined the heating rate to be $d\langle n\rangle /dt = 35 \pm 20$\,s$^{-1}$, which corresponds to an electric field noise density of {$S_{\rm{E}} = 8.0 \pm 4.6\times 10^{-13}$ \,V$^2$m$^{-2}$Hz$^{-1}$.  This heating rate is in line with heating rates for other traps of this size. More precise measurements of the heating rate would require alternative techniques such as carrier-to-sideband ratio measurements, which are not possible at this stage due to the necessary probe laser powers. The fundamental limit to the ion heating rate is expected to be the Johnson noise of the LRC circuit connected to the endcap electrodes. \cite{Brownnutt2014}.

\subsection{Thermal measurements of the trap structure}\label{sec:heat}

One of the dominant causes of uncertainty in state-of-the-art optical atomic clocks is the Stark shift caused by the emission of black-body radiation from the trap and vacuum chamber apparatus. It is therefore essential that any trap is designed with good thermal properties so that any temperature rise during operation is low, can be accurately determined,  and can be kept stable.

Finite element models (FEM) were developed by CMI \cite{Dolezal2015} to calculate the changes in the temperature distribution across the ion trap caused by its operation, and thus the effective temperature of the total BBR environment seen by the ion.    Models of the trap  show that the maximum temperature rise of any part of the trap structure should be 0.15K at standard operating conditions, and are shown in Fig.\,\ref{fig:CMIModel}.  As the solid angle subtended by the hot regions of the trap is very small, and the rf electrode, which has a very large solid angle, has been polished to a mirror finish, the effective temperature rise that should be seen by the ion is much lower than this, $\Delta T_{\left(\rm{ion}\right)} = 0.02 \pm 0.02$\,K.

\begin{figure}[htbp]
\begin{center}
\includegraphics[width=\columnwidth]{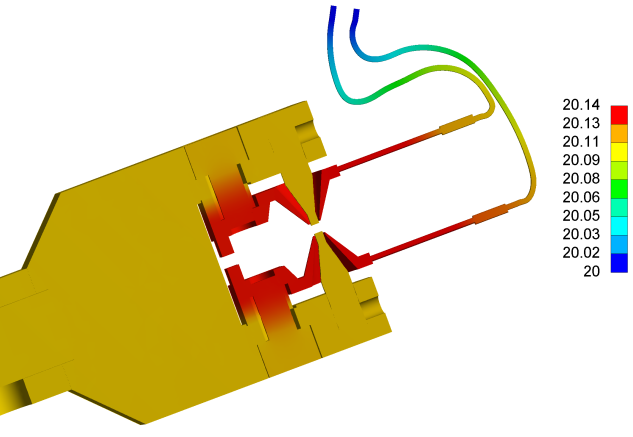}
\caption{FEM simulation of the thermal profile cross-section for an applied rf potential at $\Omega_{\rm{rf}}/2\pi=15$\,MHz required to give $\omega_{r}/2\pi=1$\,MHz . A maximum temperature rise of 0.15\,K was found in the fine wires which connect to the dc endcap electrodes.}
\label{fig:CMIModel}
\end{center}
\end{figure}

The thermal characteristics of the operating trap were measured in situ through the MgF$_{2}$ window using a Cedip Silver thermal infrared camera in the 3-5\,$\mu$m radiation band Fig.\,\ref{fig:ThermalImage}. The imaging setup was pre-calibrated by measuring target samples at known temperatures. During initial characterization of the trap \cite{Dolezal2015} it was found that dielectric heating in the UHV feedthrough completely dominated the heating from the trap structure itself, causing a temperature rise of 6.5\,K when secular frequencies of 1\,MHz were generated at a drive frequency of 21\,MHz. A simple heat sink made from a thermally conductive ceramic post (Shapal $\diameter=13$\,mm) reduced this temperature rise by a factor of two, and a further improvement in the feedthrough temperature was obtained by decreasing the drive frequency by 30\%.  With these minor modifications a temperature rise of 1.5\,K was measured both with the thermal camera and calibrated PT100 sensors on the air-side of the feedthrough, when an ion was trapped in a $\Omega_{\rm{rf}}/2\pi=14\,$MHz field with $\omega_{r}/2\pi=1\,$MHz. This results in an effective temperature rise at the ion of $\Delta T_{\left(\rm{ion}\right)} = 0.14 \pm0.14$\,K.

\begin{figure}[htbp]
\begin{center}
\includegraphics[width=\columnwidth]{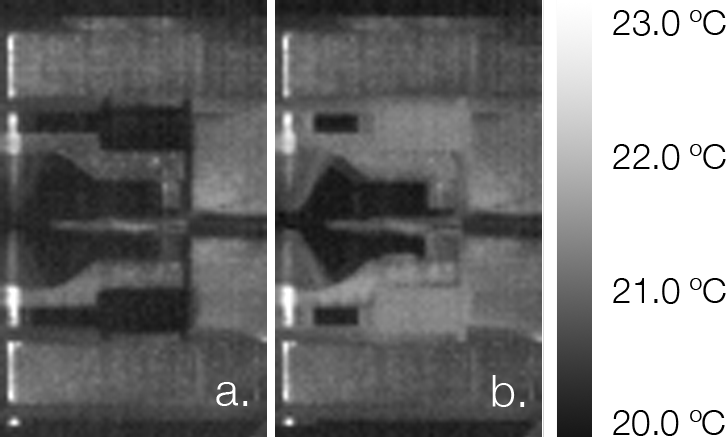}
\caption{IR images of the trap structure taken in the 3-5$\,\mu$m range (a) without an applied rf voltage, and (b) with $\Omega_{\rm{rf}}/2\pi=14\,$MHz field at $\omega_{r}/2\pi=1\,$MHz.  The IR images are not corrected for emissivity and thus show the effective temperature that the ion  observes rather than the actual temperature of the trap. Reflections belong to ambient laboratory radiation which is shielded during normal operation.}
\label{fig:ThermalImage}
\end{center}
\end{figure}

\section{Conclusion}

We have designed and constructed a trap for single ions with the endcap topology in order to minimize trap-ion interactions and thus approach an ideal isolated quantum system.   The trap has been characterized using a single $^{171}$Yb$^{+}$ ion to investigate the micromotion and anomolous heating rate properties of the trap. The trap was operated over a range of frequencies from $\Omega_{\rm{rf}}/2\pi=3 \rightarrow 21$\,MHz, and with secular frequencies $\omega_{r,z}$ from $200\,$kHz to the MHz range. Cooled ions remain in the trap almost indefinitely, and the uncooled ion-retention time is in excess of 15\,hrs. A fluorescence signal-to-noise ratio from a single ion of 400 is achieved. 

 Excess micromotion has been minimized to the level of $S_{\rm{mod}} = 0.000(5)$ as determined via the rf-photon correlation technique.  The anomalous phonon heating rate of the trap was measured using the Doppler recooling method to be $d\langle n\rangle /dt = 50 ^{+100}_{-50}$\,s$^{-1}$ and via the observation of  Rabi flop dephasing to be $d\langle n\rangle /dt = 35 \pm 20$\,s$^{-1}$ with the uncertainty limited by the experimental methods.   Further investigation of the anomalous phonon heating rate using sideband spectroscopy is planned.   The effective rise in the BBR temperature at the ion has been determined via a combination of thermal imaging and FEM models to be $\Delta T_{\left(\rm{ion}\right)}=0.14\pm 0.14$\,K limited by the rf loss in the vacuum feedthrough. 

When used for an atomic clock based on the frequency of the E3: $^{2}S_{1/2}-^{2}F_{7/2}$ transition in $^{171}$Yb$^{+}$,  perturbations induced by these trap parameters are presented in Table\,\ref{tab:ErrBud}. These values are calculated using the methods presented in \mbox{\cite{Dube2013}} with the scalar polarisability $\Delta\alpha_{\rm{S}}^{\rm{dc}}=1.3(6)\times10^{-40}$\,J\,V$^{2}$\,m$^{-2}$ \mbox{\cite{Huntemann2012}}. These contributions, when summed in quadrature, constitute a fractional uncertainty on the transition frequency of $3.5\times10^{-19}$.    This should be compared with the uncertainty contribution from trap-based frequency shifts in recently published absolute frequency measurements of $\sim1.5\times10^{-17}$ \cite{Godun2014a,Huntemann2012}. The uncertainty could be further improved by reducing the secular shifts through well established techniques for ground-state cooling of the ion, and by fully heat-sinking the feedthrough.  If successful, these improvements should reduce the uncertainty in the error budget by more than an order of magnitude.\\

\begin{table}
\begin{tabular}{lrr}
\toprule
\bfseries{Source of Shift} & $\delta\nu/\nu_{0}(10^{-18})$ & $\sigma/\nu_{0}(10^{-18})$\\
\midrule
Background gas collisions 			& 0.15	& 0.15\\
Excess $\mu$-motion scalar dc-Stark  	& 0 		& 0.1 \\
Excess $\mu$-motion time dilation  	& 0 		& 0.1 \\
Secular motion scalar dc-Stark  		& -0.3	& 0.2\\
Secular motion time dilation  		& -0.7 		& 0.1\\
BBR dc-Stark: offset from 293\,K				 	& -0.2 		& 0.2 \\
\bottomrule
\bfseries{TOTAL} & -1.05  & 0.35 \\
\bottomrule
\end{tabular}
\caption{Error budget for trap induced shifts and uncertainties for the E3: $^{2}$S$_{1/2}-^{2}$F$_{7/2}$ transition in $^{171}$Yb$^{+}$ when the trap is operated at $\Omega_{\rm{\rm{rf}}}/2\pi=14$\,MHz, with $\omega_{x,y,z}/2\pi=(1,1,2)$\,MHz.  }
\label{tab:ErrBud}
\end{table}

This work was funded by the European Metrology Research Programme (EMRP), the UK National Measurement System, and the European Space Agency (ESA). The EMRP is jointly funded by the EMRP participating countries within EURAMET and the European Union.

\bibliographystyle{ieeetr}

\end{document}